\documentclass[10pt,aps,pra,amsmath,amssymb,amsfonts,floatfix,twocolumn,superscriptaddress]{revtex4-2}

\usepackage[utf8]{inputenc} 
\usepackage[T1]{fontenc}
\usepackage{graphicx}
\usepackage{bm}    
\usepackage{amsbsy}
\usepackage[separate-uncertainty]{siunitx}
\usepackage[breaklinks=true]{hyperref}
\usepackage{hyperref}
\usepackage{booktabs}
\hypersetup{
	colorlinks=true,
	linkcolor=blue,
	citecolor=blue,
	urlcolor=blue
}

\begin{document}
	
\title{Effect of interparticle fields and radiation reaction on beam dynamics}

\author{Michael~J.~Quin}
\email{michael.quin@physics.gu.se}
\affiliation{Max-Planck-Institut f\"{u}r Kernphysik, Saupfercheckweg 1, 69117 Heidelberg, Germany}

\author{Antonino~Di~Piazza}
\email{a.dipiazza@rochester.edu}
\affiliation{Max-Planck-Institut f\"{u}r Kernphysik, Saupfercheckweg 1, 69117 Heidelberg, Germany}
\affiliation{Department of Physics and Astronomy, University of Rochester, Rochester, NY 14627, USA}
\affiliation{Laboratory for Laser Energetics, University of Rochester, Rochester, NY 14623, USA}

\author{Christoph~H.~Keitel}
\affiliation{Max-Planck-Institut f\"{u}r Kernphysik, Saupfercheckweg 1, 69117 Heidelberg, Germany}

\author{Matteo~Tamburini}
\email{matteo.tamburini@mpi-hd.mpg.de}
\affiliation{Max-Planck-Institut f\"{u}r Kernphysik, Saupfercheckweg 1, 69117 Heidelberg, Germany}

\date{\today}

\begin{abstract}
\noindent
The dynamics of relativistic particles in an intense electromagnetic field can be described by the Landau-Lifshitz (LL) equation, where the adiation reaction (RR) is accounted for via a self-force, and interparticle fields are often neglected as an approximation. However, the inclusion of interparticle fields is necessary to ensure energy-momentum conservation, particularly during coherent emission. Here we present (i) an analytical proof showing that the energy-momentum conservation law of the Hamilton-Rohrlich-Dirac action, which is divergence free and describes a generic system of interacting charges, respects causality and provides physically sensible results; (ii) a simple generalization of the LL equation for many particles evaluated as a function of the total field, i.e., the sum of the external and interparticle fields. By performing first-principles numerical simulations of a neutral, relativistic bunch of electrons and positrons ($e^-/e^+$) colliding with a laser pulse, this theory is shown to satisfy energy-momentum conservation when interparticle fields and RR are simultaneously taken into account; and (iii) the combined effect of interparticle fields and RR primarily affects the tail of the particle energy distribution. Additionally, our first-principles simulations show that the effect of interparticle fields on beam energy loss becomes smaller when most of the radiated energy is incoherent.
\end{abstract}
 
\maketitle

\section{Introduction}

\noindent
A point particle of charge $e$ and mass $m$, where $e=-|e|$ for an electron, emits electromagnetic radiation when accelerated. Yet the Lorentz equation, featuring only the external electromagnetic accelerating force, neglects this emission when describing the trajectory, apparently violating energy-momentum conservation. The radiation reaction (RR) problem then refers to how one should incorporate the particle's self-field, which diverges on the world line, to create a self-consistent equation of motion. Typically, the divergent part of the self-field is
removed by mass renormalization, by which one derives the Lorentz-Abraham-Dirac (LAD) equation~\cite{dirac1938, abraham1905, teitelboim1970, hamilton1971, rohrlich1964, rohrlich2007, barut1980}. However, the LAD equation permits unphysical ``runaway'' solutions~\cite{rohrlich2001}. This problem can be resolved by a perturbative expansion of the self-force by which one obtains the Landau-Lifshitz (LL) equation~\cite{landaulifshitz_vol2}.

The LAD and LL equations are often solved analytically for a single particle in an \textit{external field}, such as a constant magnetic field~\cite{borisov1972}, plane wave~\cite{dipiazza2008}, or the field of a crystal~\cite{dipiazza2017}. These solutions are often used to predict the energy lost by a beam of ultrarelativistic electrons propagating through the field of a laser pulse~\cite{cole2018, poder2018,los2024} or aligned crystal~\cite{nielsen2021}, implicitly assuming that \textit{interparticle fields} can be neglected as an approximation. To date, these experiments provide the most convincing evidence of the LL equation's validity. However, the Poynting vector varies quadratically with the total field, and so this approach of treating particles independently is inconsistent with energy-momentum conservation as a matter of principle~\cite{gromes2015, abraham1905}. Therefore, it is interesting to explore scenarios in which the self-force, coherent emission and interparticle fields play an important role. 

Note that the LAD self-force is proportional to the classical electron radius $r_e=e^2/m\approx2.8\,\si{\femto\metre}$, where we employ natural units $c=\hbar=4\pi\varepsilon_0=1$. This has led some authors to suggest that a `small' bunch of $N$ electrons could be modeled as a single point particle of charge $Ne$ and mass $Nm$, which would experience a `coherently enhanced' self-force proportional to $Nr_e$~\cite{kaplan2002, smorenburg2010}. Here, we interpret `small' with respect to the characteristic wavelength of emitted radiation in a given external field. 

It is unclear how this model would apply to a neutral, relativistic bunch of electrons and positrons ($e^-/e^+$). At first glance, one might model an $e^-/e^+$ bunch as a neutral point particle which emits no radiation and experiences no self-force. Alternatively, if the $e^-/e^+$ bunch was polarized by an external electric field, we could identify an \textit{ad hoc} self-force from the emission of dipole radiation~\cite[\S\,75]{landaulifshitz_vol2}. 
A major difficulty with these ad hoc models is their assumption that the interparticle fields can be neglected. This leads to a uniform loss of energy caused by evaluating the self-force with the external field alone. However, relativistic particles radiate anisotropically, such that, once interparticle fields are taken into account, each particle experiences a distinct local field. Consequently, both the Lorentz force and the self-force vary from particle to particle.

Various theories attempt to describe the trajectories of point particles in a self-consistent manner. One approach assigns a fraction of the power radiated to each particle \textit{ad hoc}~\cite{kimel1995}, which accounts for coherent emission but neglects work done by Coulomb-like velocity fields. Other models ignore the divergent self-interaction of a particle with its own field, eliminating RR entirely while preserving energy-momentum conservation~\cite{gratus2022}, but it is difficult to reconcile this approach with the experimental evidence for RR. An ideal theory would agree with experimental observations, respect causality, and preserve energy-momentum conservation.

In this paper, we review the equations of motion and conservation laws which result from the Hamilton-Rohrlich-Dirac action~\cite{hamilton1971}, and present a simple generalization of the LL equation for many point particles. We show that this theory is divergence free, respects causality, and satisfies energy-momentum conservation as expected. This theory is applied in simulations of an $e^-/e^+$ bunch colliding with a laser pulse, in a regime where the external field dominates over the interparticle fields. Here we show that the interparticle fields affect the particle dynamics primarily through the Lorentz force, and the self-force provides an important correction needed to ensure energy-momentum conservation.

\section{Radiation reaction with many bodies}

\label{sec:HRD_model}

\noindent 
Throughout this paper we employ the Minkowski metric $\eta_{\mu\nu} = \text{diag}(+1, -1, -1, -1)$ and shorthand notation for the inner product $(ab) = a^\mu b_\mu$, square $(a)^2=(aa)$, and antisymmetric part $a^{[\mu} b^{\nu]} = \frac{1}{2}(a^\mu b^\nu - a^\nu b^\mu)$. Note that every field is the derivative of a potential with the same notation. For example, we can construct the external field $F^{\mu\nu}_{\text{ext}}(x)= 2\, \partial^{[\mu}_{\phantom{\text{A}}} A^{\nu]}_{\text{ext}}(x)$ from the external potential $A^\mu_{\text{ext}}(x)$.

\subsection{Action free of divergences}

\noindent
Consider $N$ particles, each of charge $e_i$ and mass $m_i$, in an external electromagnetic field $F^{\mu\nu}_{\text{ext}}\equiv F^{\mu\nu}_{\text{ext}}(x)$. The action for this system consists of kinetic, interaction and field terms~~\cite[Eq.~(27.7)]{landaulifshitz_vol2}
%
% generic, many-particle action e.g. LL Sec. 27
\begin{equation}
	\begin{split}
		% kinetic and interaction terms
		S = &-\sum^N_{i=1} \int \Big[ m_i\,\sqrt{u^2_i\,} + e_i A^\mu(x_i)\,u_{i,\mu} \Big]\,d\tau_i 
		\\
		% fields
		&-\frac{1}{16\pi} \int F^{\mu\nu} F_{\mu\nu}\,d^4x.
	\end{split}
	\label{eq:S_generic}
\end{equation}
Here we define the four-vector position $x^\mu_i\equiv x^\mu_i(\tau_i)$ and velocity $u^\mu_i\equiv u^\mu_i(\tau_i)=\dot{x}^\mu_i$ of particle $i$, where the dot denotes a derivative with respect to the proper time $\tau_i$ \footnote{Rigorously speaking in Eq. (\ref{eq:S_generic}) one should use for each integral a monotonic function $s_i(\tau_i)$ of the proper time as integration variable and only after the equations of motion are derived identify $s_i(\tau_i)=\tau_i$.}. The separation $R^\mu_i \equiv R^\mu_i(\tau_i)$ of the observer and source can be written as $R^\mu_i(\tau_i)=x^\mu-x^\mu_i(\tau_i)$.

The solution of Maxwell's equations for a pointlike source can be found by the method of Green's functions~\cite{rohrlich2007,jackson1998}, and are known as the retarded and advanced potentials
\begin{equation}
    A^\mu_{\substack{\text{ret}\,i\\\text{adv}\,i}}(x) = \pm \Bigg[\frac{e_i u^\mu_i}{(R_i u_i)}\Bigg]_{\substack{\tau_{\text{ret}\,i}\\ \tau_{\text{adv}\,i}}}.
    \label{eq:Aretadv}
\end{equation}
These are evaluated at the retarded $\tau_i=\tau_{\text{ret}\,i}$ and advanced proper time $\tau_i=\tau_{\text{adv}\,i}$ respectively, which satisfy the null condition $(R_i(\tau_{\text{ret}\,i}))^2=(R_i(\tau_{\text{adv}\,i}))^2=0$. Here we have introduced a minus sign instead of applying the modulus in the advanced case $|(R_iu_i)_{\tau_{\text{adv\,i}}}| = -(R_iu_i)_{\tau_{\text{adv\,i}}}$. Note that the retarded potential is observed after emission $R^{0}_i(\tau_{\text{ret}\,i})>0$, and is said to respect causality, while the advanced potential is observed before emission $R^{0}_i(\tau_{\text{adv}\,i})<0$, and is said to violate causality. By differentiating these potentials, one obtains the retarded and advanced fields~\cite{rohrlich2007,jackson1998}
\begin{equation}
    \begin{split}
        F^{\mu\nu}_{\substack{\text{ret}\,i\\\text{adv}\,i}} (x) = \pm\Bigg[\Bigg.&\frac{2e_iR_i^{[\mu}u_i^{\nu]}}{(R_iu_i)^3} + \frac{2e_iR_i^{[\mu}\mathrm{a}_i^{\nu]}}{(R_iu_i)^2} 
	\\
        &-\frac{2e_iR_i^{[\mu}u_i^{\nu]}}{(R_iu_i)^3}(R_i\mathrm{a}_i) \Bigg.\Bigg]_{\substack{\tau_{\text{ret}\,i} \\ \tau_{\text{adv}\,i} }},
    \end{split}
    \label{eq:Fretadv}
\end{equation}
where $\mathrm{a}^\mu_i\equiv\dot{u}^\mu_i=du^\mu_i/d\tau_i$ is the four-acceleration. It will prove useful to introduce the plus and minus potentials
\begin{align}
    % A plus/minus
    A^\mu_\pm(x) &\equiv \frac{1}{2} \left[ A^\mu_{\text{ret}}(x) \pm A^\mu_{\text{adv}}(x) \right] = \sum^N_{i=1} A^\mu_{\pm\,i}(x) \nonumber
    \\
    &= \frac{1}{2} \sum^N_{i=1} \left[ A^\mu_{\text{ret}\,i}(x) \pm A^\mu_{\text{adv}\,i}(x) \right],
    \label{eq:A_plusminus}
\end{align}
which can be interpreted by expanding their corresponding fields close to the charge. One can show to leading order in $\mathcal{R}_i(\tau_i)=|\bm{x}-\bm{x}_i(\tau_i)|$ that the plus and minus fields are~\footnote{The leading order expression for the minus field in Eq.~\eqref{eq:F_minus} can be found in Ref.~\cite[Eq.~(6-63)]{rohrlich2007}, while the plus field in Eq.~\eqref{eq:F_plus}] can be derived from the retarded and advanced fields in Ref.~\cite[Eq.~(6-62)]{rohrlich2007}. Note that this reference uses a different metric $\text{diag}(-1, +1, +1, +1)$ compared to this paper.}
\begin{align}
    F^{\mu\nu}_{+i}(x) &\approx \frac{2e_i}{(R_iu_i)^3}\,R^{[\mu}_i u^{\nu]}_i,
    \label{eq:F_plus}
    \\
    F^{\mu\nu}_{-\,i}(x) &\approx \frac{4e_i}{3}\,\dot{\mathrm{a}}^{[\mu}_i u^{\nu]}_i.
    \label{eq:F_minus}
\end{align}
Close to the particle in its instantaneous rest frame, the plus field is a Coulomb field which diverges in the limit $\mathcal{R}_i\rightarrow0$ and so we interpret it as being `bound' to the charge. This assertion is investigated in detail by Teitelboim~\cite{teitelboim1970}. Meanwhile, the minus field is finite everywhere and satisfies the free (sourceless) Maxwell's equations, and so it is said to be `unbound'. It is the minus field which Dirac associates with radiation reaction~\cite{dirac1938}.

By defining the total potential as the sum of the external and retarded potential,
\begin{align}
    A^\mu(x) &= A^\mu_{\text{ext}}(x) + A^\mu_{\text{ret}}(x),
    \label{eq:Atot}
    \\
    A^\mu_{\text{ret}}(x) &= \sum^N_{i=1} A^\mu_{\text{ret}\,i}(x),
\end{align}
we can see that the retarded potential is the sum of $A^\mu_-(x)$, which is finite everywhere, and a bound part $A^\mu_+(x)$, which contains divergences. Therefore, we are motivated to separate the total potential into free $A^\mu_f(x)$ and bound $A^\mu_+(x)$ parts:
\begin{align}
    % A total: free and bound parts
    A^\mu(x) &= A^\mu_f(x) + A^\mu_+(x),
    \label{eq:Afreebound}
    \\
    % A free
    A^\mu_f(x) &= A^\mu_{\text{ext}}(x) + A^\mu_-(x).
    \label{eq:Afree}
\end{align}
Similarly, we can separate the total field into free and bound parts $F^{\mu\nu}(x)=F^{\mu\nu}_f(x)+F^{\mu\nu}_+(x)$. This separation can also be applied to the action in Eq.~\eqref{eq:S_generic}. With the four-dimensional volume element $d^4x=\mathcal{R}^2d\mathcal{R}d\Omega dt$, where $d\Omega$ is the solid angle element, we therefore recognize that the following terms in the action are divergent
\begin{equation}
    -\sum^N_{i=1} \Bigg[\Bigg. e_i \int A^\mu_{+i}(x_i)\,u_{i,\mu}\,d\tau_i +\frac{1}{16\pi} \int F^{\mu\nu}_{+i} F^{}_{+i,\mu\nu}\,d^4x \Bigg.\Bigg].
    \label{eq:divergent_terms}
\end{equation}
One argues that these terms are associated with energy bound to the particles and should therefore be discarded~\cite{hamilton1971} or removed by a classical mass renormalization procedure~\cite{rohrlich2007,barut1980}.
One then obtains what we refer to as the Hamilton-Rohrlich-Dirac (HRD) action~\cite{hamilton1971}
\begin{equation}
    S = \sum^N_{i=1}\int L_i\,d\tau_i + \int\mathcal{L}_F\,d^4x,
    \label{eq:Srohrlichhamilton}
\end{equation}
written in terms of the Lagrangian for each particle  
\begin{equation}
    L_i = -m_i\sqrt{u_i^2\,} - e_i A^\mu_f(x_i)\,u_{i,\mu} - e_i \sum^N_{\substack{j=1\\j\neq i}}  A^\mu_{+j}(x_i)\,u_{i,\mu},
    \label{eq:Lparticles}
\end{equation}
and the Lagrangian density for the fields
\begin{equation}    
    \mathcal{L}_F = -\frac{1}{16\pi} \Big[ F^{\mu\nu}_f F^{\,}_{f,\mu\nu} + 2 F^{\mu\nu}_f F^{\,}_{+,\mu\nu} + \sum^N_{\substack{i,\,j=1\\j\neq i}} F^{\mu\nu}_{+i} F^{}_{+j,\mu\nu} \Big].
    \label{eq:Lfields}
\end{equation}
This action was derived by Hamilton~\cite{hamilton1971} who closely follows the work of Rohrlich~\cite{rohrlich1964, rohrlich2007}. However, Rohrlich's derivation omits the last term of $\mathcal{L}_F$, which vanishes in the case of a single particle, in the belief that it is divergent, though it is not. The Lagrangian $L_i$ was also obtained by Dirac~\cite[Eq.~(45)]{dirac1938}. Note that the HRD action does not depend on the derivatives of particle velocity and is free from divergences, provided that two particles never occupy \textit{exactly} the same point in spacetime, which cannot occur in the classical regime. According to Ref.~\cite{hamilton1971}, the independent quantities in the action $S$ are the position of the particles as well as the fields $A_{\text{ext}}^{\mu}(x)$, $A_{-\,i}^{\mu}(x)$ and $A_{+\,i}^{\mu}(x)$, with the variations with respect to $A_{\text{ext}}^{\mu}(x)$ and $A_{-\,i}^{\mu}(x)$ giving rise to the same equations (see Ref.~\cite{hamilton1971}).

\subsection{Equations of motion}

\noindent
With the action defined, we can now describe the dynamics of the particles. As usual, we vary the action with respect to the trajectory of particle $i$ while holding the fields constant, and then we apply the principle of least action. This procedure leads to the Euler-Lagrange equation
\begin{equation}
    \frac{d}{d\tau_i}\left(\frac{\partial L_i}{\partial u_{i,\mu}}\right) = \frac{\partial L_i}{\partial x_{i,\mu}},
    \label{eq:EulerLag}
\end{equation}
from which we can obtain the LAD equation of motion~(see Appendix~\ref{ap:LAD} for a summary of the steps leading from Eq.~\eqref{eq:Lparticles} to Eq.~\eqref{eq:LADeq}, where $L_i$ is independent of derivatives of the particle's velocity)
\begin{equation}
    m_i\mathrm{a}_i^\mu = e_i \mathcal{F}^{\mu\nu}_i u_{i,\nu} + \frac{2}{3} e_i^2 \left( \dot{\mathrm{a}}^\mu_i + \mathrm{a}^2_i u^\mu_i \right).
    \label{eq:LADeq}
\end{equation}
Note that this equation was also obtained by Dirac~\cite[Eq.~(41)]{dirac1938}. To describe the trajectory, we require the total field excluding the self-field, defined as
\begin{equation}
	\begin{split}
		\mathcal{F}^{\mu\nu}_i(x) &= F^{\mu\nu}(x) - F^{\mu\nu}_{\text{ret}\,i}(x)
		\\%[0.1em]
		&= F^{\mu\nu}_{\text{ext}}(x) + \sum^N_{\substack{j=1\\j\neq i}} F^{\mu\nu}_{\text{ret}\,j}(x).
		\label{eq:Ftot}
	\end{split}
\end{equation}
This field is then evaluated at the location of the particle $\mathcal{F}^{\mu\nu}_i\equiv\mathcal{F}^{\mu\nu}_i(x_i)$ and does not depend on the causality-violating advanced fields. When the retarded field $F^{\mu\nu}_{\text{ret}\,j}(x_i)$ from particle $j$ is observed by a particle $i\neq j$, we refer to it as an \textit{interparticle field}.

However, the LAD equation admits unphysical solutions which violate energy-momentum conservation~\cite{rohrlich2001}. This problem can be avoided by performing a perturbative expansion of the self-force as described by Landau \& Lifshitz~\cite[\S\,76]{landaulifshitz_vol2}. Following their method, we approximate the four-acceleration in the LAD self-force with the Lorentz equation ${m_i\mathrm{a}_i^\mu \approx e_i \mathcal{F}^{\mu\nu}_i u_{i,\nu}}$. This is automatically consistent in the classical regime where quantum effects can be neglected. The resulting LL equation is 
\begin{align}
    m_i\mathrm{a}^\mu_i =&~e_i \mathcal{F}^{\mu\nu}_i u_{\nu,i} + \frac{2e^3_i}{3m_i} \left(\partial_{i,\alpha} \mathcal{F}^{\mu\nu}_i \right) u^{\alpha}_i u_{i,\nu} \nonumber
    \\
    &+ \frac{2e^4_i}{3m_i^2} \Big[\Big. \mathcal{F}^{\mu\nu}_i \mathcal{F}_{\nu\alpha,i} u^\alpha_i + (\mathcal{F}_iu_i)^2 u^\mu_i \Big.\Big].
    \label{eq:LLeq}
\end{align}
Note the presence of the interparticle fields in the LL self-force. 
If the interparticle fields add coherently and become comparable to the external field, then this will drastically alter the value of the self-force.
Although this is not the regime which we simulate in Sec.~\ref{sec:simulations}, we show that, even in this regime, the inclusion of both interparticle fields and RR can be critical to ensure energy-momentum conservation. Note that the interplay of the Lorentz force and self-force can lead to a qualitatively different dynamics as demonstrated in Ref.~\cite{tamburini2014}. 
This is distinct from the `coherently enhanced’ self-force used in previous models~\cite{kaplan2002, smorenburg2010}, which omit the interparticle fields and instead assume \textit{a priori} that the self-force scales proportionally with the number of particles (see also the introduction).

Both the LAD and LL equations are valid only in the regime of classical electrodynamics. Therefore, the instantaneous rest-frame field must be small compared to the critical field $F_{\text{cr}}=m^2/|e|\approx1.3\times10^{18}\,\si{\volt/\metre}$ of quantum electrodynamics, that is ${\chi_i(x_i)=\sqrt{|(\mathcal{F}_i(x_i)u_i)^2|}/F_{\text{cr}}\ll1}$~\cite{dipiazza2012_review,gonoskov2022,fedotov2023}. In an external laser field, as is considered here, one often argues that the derivatives of the fields are negligible within the classical regime~\cite{tamburini2010, tamburini2011_phd}. By discarding these terms, one obtains the reduced LL equation 
\begin{equation}
    m_i\mathrm{a}^\mu_i = e_i \mathcal{F}^{\mu\nu}_i u_{\nu,i} + \frac{2e^4_i}{3m_i^2} \Big[\Big. \mathcal{F}^{\mu\nu}_i \mathcal{F}_{\nu\alpha,i} u^\alpha_i + (\mathcal{F}_iu_i)^2 u^\mu_i \Big.\Big].
    \label{eq:redLLeq}
\end{equation}
As we shall see, in our simulations the interparticle fields are small compared to the external field, and hence this approximation is applicable here. In practice, it is the reduced LL equation which is solved by our code, and from now on we will simply refer to it as the LL (or `Lan-Lif') equation.

\subsection{Energy-momentum conservation}

\noindent
From the HRD action, we can define a series of conservation laws according to Noether's theorem. We will restrict our attention to conservation of energy and momentum. The change in momentum of the particles is related to the momentum carried by the electromagnetic fields by~\cite{hamilton1971}
\begin{equation}
	\partial_\nu T^{\mu\nu}_{p} + \partial_\nu T^{\mu\nu}_{\text{HRD}} = 0.
		\label{eq:P_consv2}
\end{equation}
Here we have introduced the energy-momentum tensor for the particles $T^{\mu \nu}_{p}\equiv T^{\mu \nu}_{p}(x)$. The corresponding mechanical momentum is
\begin{equation}
	P^\mu_p = \int T^{\mu 0}_{p} d^3 x = \sum^N_{i=1} m_i u^\mu_i,
\end{equation}
while the energy-momentum tensor associated with the fields is
\begin{align}
	T^{\mu\nu}_{\text{HRD}} \equiv &~T^{\mu\nu}(F_f, F_f) + T^{\mu\nu}(F_f, F_+) \nonumber
	\\ 
	&+ T^{\mu\nu}(F_+, F_f) + \sum^N_{\substack{i,j=1\\j\neq i}} T^{\mu\nu}(F_{+\,i}, F_{+\,j}),
\end{align}
which can be rewritten as
\begin{align}
	T^{\mu\nu}_{\text{HRD}} = &~T^{\mu\nu}(F_{\text{ext}}, F_{\text{ext}}) + T^{\mu\nu}(F_{\text{ext}}, F_{\text{ret}}) \nonumber
	\\ 
	& + T^{\mu\nu}(F_{\text{ret}}, F_{\text{ext}}) + \sum^N_{\substack{i,j=1\\j\neq i}} T^{\mu\nu}(F_{\text{ret}\,i}, F_{\text{ret}\,j}) \nonumber
	\\ 
	& + \sum^N_{i=1} \bigg[\bigg. T^{\mu\nu}(F_{-\,i}, F_{-\,i}) + T^{\mu\nu}(F_{-\,i}, F_{+\,i}) \nonumber
	\\
	& \hspace{3.5em} + T^{\mu\nu}(F_{+\,i}, F_{-\,i}) \bigg.\bigg].
       \label{eq:T_HRD}
\end{align}
Here the symmetric energy-momentum tensor is defined as a function of the fields
\begin{equation}
    T^{\mu\nu}(a, b) = \frac{1}{4\pi} \left[ a^{\mu\alpha} \eta_{\alpha\beta} b^{\beta\nu} + \frac{1}{4} \eta^{\mu\nu} a^{\alpha\beta} b^{}_{\alpha\beta} \right].
    \label{eq:T_definition}
\end{equation}

By taking the integral of the conservation law in Eq.~\eqref{eq:P_consv2} over the four-dimensional spacetime volume, and by using the divergence theorem to reduce the four-dimensional integral to the flux through a three dimensional closed hypersurface containing all particles, we obtain
\begin{equation}
    \Delta P^{\mu} = - \int_{\Sigma} T^{\mu \nu}_{\text{HRD}} \, d^3\sigma_\nu,
    \label{eq:consv_subtract}
\end{equation}
where we have introduced the total four-momentum of the system
\begin{equation}
\label{P^mu}
    P^{\mu} \equiv P^{\mu}_p + \int T^{\mu 0}_{\text{HRD}} \, d^3 x.
\end{equation}
Here $\Sigma$ is the lateral part of a hypercylinder with temporal domain $t\in[-\infty, +\infty]$ and surface element $d^3\sigma_\nu = \tilde{n}_\nu \mathcal{R}^2 d\Omega dt$, where the normal four-vector $\tilde{n}^\mu=(0, \bm{n})$ satisfies $(\tilde{n})^2=-1$. The bases of the hypercylinder correspond to the configuration space in the remote past and future. Consequently, the integrals over these bases [see the second term in Eq.~\eqref{P^mu}] represent the four-momentum of the electromagnetic field in these distant temporal regions. 

Equation~\eqref{eq:consv_subtract} already indicates that the change in the total four-momentum of the system is related to the time-integrated flux of electromagnetic energy escaping through a two-dimensional surface enclosing the system of charges. We now observe that the tensor $T^{\mu \nu}_{\text{HRD}}$ does not contain the square of the plus-field of the charges, ensuring that the conservation law in Eq.~\eqref{eq:consv_subtract} is free of divergences. Moreover, under the reasonable assumption that the particles in the remote past and future move freely with constant velocities, it can be shown that the advanced field of each particle appearing in the four-momentum of the field [see Eq.~\eqref{P^mu}] coincides with the corresponding retarded field, such that the fields $F^{\mu\nu}_{- i}$ vanish in those regions and $T^{\mu \nu}_{\text{HRD}}$ is manifestly causal [see Eq.~\eqref{eq:T_HRD}]. 

The equality of the retarded and advanced fields for constant velocity motion is apparent in the particle’s rest frame, and holds in all inertial frames by virtue of Lorentz invariance. In fact, it is well known that the retarded field coincides with the Lorentz boosted Coulomb field for a particle with constant velocity~\cite{jackson1998}. Additionally, recalling our definition of the total field from Eq.~\eqref{eq:Atot}, it is straightforward to show that  
\begin{equation}
    T^{\mu\nu}_{\text{HRD}} = T^{\mu\nu}(F, F) - \sum^N_{i=1} T^{\mu\nu}(F_{+\,i}, F_{+\,i}).
    \label{eq:Ttotalminusdiv}
\end{equation}
This expression seems to suggest that the right-hand side of Eq.~\eqref{eq:consv_subtract} still depends on advanced fields. However, we intuitively expect that the change in the total four-momentum of the particles and the field should be equal to the time-integrated flux of the escaping electromagnetic field, involving only the external and retarded fields, i.e., computed from the total field \( F^{\mu\nu}(x) \). In other words, we anticipate that  
\begin{equation}
    \int_\Sigma T^{\mu\nu}(F_{+\,i}, F_{+\,i}) \, d^3\sigma_\nu = 0,
    \label{eq:Tplus_zero}
\end{equation}
which simplifies the conservation law in Eq.~\eqref{eq:consv_subtract} to  
\begin{equation}
    \Delta P^{\mu} = - \int_{\Sigma} T^{\mu\nu}(F, F) \, d^3\sigma_\nu.
    \label{eq:Ttotalconserv_complete}
\end{equation}
Below, we will prove Eq.~\eqref{eq:Tplus_zero}, leading to the conclusion that the conservation law in the HRD model is not only free of divergences but also respects causality, as it does not depend on advanced fields. 

For the physical system under investigation here, the energy-momentum conservation law in Eq.~\eqref{eq:Ttotalconserv_complete} can be further simplified. Indeed, for a neutral and nearly copropagating bunch of charges, as considered here, the contribution of the electromagnetic field of the particle to the total four-momentum of the system provides a negligible contribution compared to the total four-momentum of the charges. Since the external field is unchanged by assumption, we have that  $\Delta P^{\mu} \approx \Delta P^{\mu}_p$ and then
\begin{equation}
    \Delta P^{\mu}_p \approx - \int_{\Sigma} T^{\mu\nu}(F, F) \, d^3\sigma_\nu.
    \label{eq:Ttotalconserv}
\end{equation}

We will now prove Eq.~\eqref{eq:Tplus_zero}. We can write the change in momentum associated with the plus field as
\begin{equation}
    \frac{dP^\mu_+}{d\Omega} = \sum^N_{i=1} \int^{+\infty}_{-\infty} T^{\mu\nu}(F_{+i}, F_{+i}) \tilde{n}_\nu \mathcal{R}^2 dt.
    \label{eq:dP+dOmega}
\end{equation}
To evaluate this integral we require expressions for the fields. For a distant observer, the source-observer separation becomes approximately constant in time and equal for all particles, that is $R^\mu_i(\tau_{\text{ret}\,i}) \approx \mathcal{R} n^\mu_+$ in the retarded case and $R^\mu_i(\tau_{\text{adv}\,i}) \approx \mathcal{R} n^\mu_-$ in the advanced case. Here $n^\mu_\pm=(\pm1,\bm{n})$ are null four-vectors which satisfy $(n_\pm)^2=0$ and $(n_-n_+)=-2$. Therefore, in the limit $\mathcal{R}\rightarrow+\infty$, we can approximate the retarded and advanced fields in Eq.~\eqref{eq:Fretadv} as
\begin{align}
    F^{\mu\nu}_{\substack{\text{ret}\,i\\\text{adv}\,i}}(x) &\approx \pm\Bigg[\Bigg. \frac{2e_i}{\mathcal{R} (n_\pm u_i)} \frac{d}{d\tau_i} \left( \frac{n_\pm^{[\mu} u_i^{\nu]}}{(n_\pm u_i)} \right) \Bigg.\Bigg]_{\substack{\tau_{\text{ret}\,i} \\ \tau_{\text{adv}\,i} }}.
    %\nonumber
    \label{eq:Fretadv_asymptotic}
\end{align}
This is equivalent to retaining only the acceleration-dependent fields, which represent emitted radiation. 
With the fields known, we can evaluate the energy-momentum tensor. This process is greatly simplified by the fact that we only require the projection of the energy-momentum tensor onto the normal, that is $T^{\mu\nu}(F_{+i}, F_{+i}) \tilde{n}_\nu$. Notice that the normal can be conveniently written as $\tilde{n}^\mu=\frac{1}{2}(n^\mu_+ + n^\mu_-)$ and satisfies $(\tilde{n}n_\pm)=-1$. One can show that the cross terms involving the retarded and advanced field vanish identically even before integration~(see Appendix~\ref{ap:Tasymptotic})
\begin{equation}
    T^{\mu\nu}(F_{\text{ret}\,i}, F_{\text{adv}\,i})\tilde{n}_\nu + 
    T^{\mu\nu}(F_{\text{adv}\,i}, F_{\text{ret}\,i})\tilde{n}_\nu = 0.
    \label{eq:Tcross_zero}
\end{equation}
Therefore, we only need to compute terms which involve either the retarded or advanced field
\begin{align}
    T^{\mu\nu}(&F_{+i}, F_{+i})\tilde{n}_\nu = \nonumber
    \\
    & \frac{1}{4}\Big[T^{\mu\nu}(F_{\text{ret}\,i}, F_{\text{ret}\,i}) + 
    T^{\mu\nu}(F_{\text{adv}\,i}, F_{\text{adv}\,i}) \Big]\tilde{n}_\nu.
\end{align}
These terms can be expressed as follows~(see Appendix~\ref{ap:Tasymptotic})
\begin{align}
    T^{\mu\nu}\Big( F_{\substack{\text{ret}\,i\\\text{adv}\,i}}, &F_{\substack{\text{ret}\,i\\\text{adv}\,i}} \Big) \tilde{n}_\nu = \frac{1}{4\pi} \rho_{\pm i}\left(\tau_{\substack{\text{ret}\,i\\ \text{adv}\,i}}\right) n^\mu_\pm,
    \label{eq:Tret2}
    \\
    \rho_{\pm i}(\tau_i) &= 
    \left(\frac{e_i}{\mathcal{R}(n_\pm u_i)} \frac{d}{d\tau_i} \left[\frac{u^\mu_i}{(n_\pm u_i)} \right] \right)^2.
    \label{eq:hpm}
\end{align}

To explain how the integration over the solid angle is carried out, it will be convenient to use three-vector notation $P^\mu_+=(\mathcal{E}_+, \bm{P}_+)$ and $u^\mu_i=(\gamma_i, \bm{u}_i)$. Therefore, we can express the components of the plus-momentum as
\begin{align}
    \frac{d\mathcal{E}_+}{d\Omega} &= \frac{1}{16\pi}\sum^N_{i=1} \int^{+\infty}_{-\infty} \Big[ f_i\left(\bm{n}, \bm{u}_i\right) - f_i\left(\bm{n}, -\bm{u}_i\right) \Big]  d\tau_i,
    \label{eq:E+}
    \\ 
    \frac{d\bm{P}_+}{d\Omega} &= \frac{1}{16\pi}\sum^N_{i=1} \int^{+\infty}_{-\infty} \Big[ f_i\left(\bm{n}, \bm{u}_i\right) + f_i\left(\bm{n}, -\bm{u}_i\right) \Big] \bm{n} \,d\tau_i. \label{eq:dPdomega}
\end{align}
Note that we have changed the variable of integration to a dummy variable $\tau_i$ via $dt \approx (n_+u_i(\tau_{\text{ret}\,i}))d\tau_{\text{ret}\,i}$ in the retarded case and $dt \approx -(n_-u_i(\tau_{\text{adv}\,i}))d\tau_{\text{adv}\,i}$ in the advanced case. Finally, the function in the integrand can be written as~(see Appendix~\ref{ap:threevector})
\begin{equation}
    f_i\left(\bm{n}, \bm{u}_i\right) = \frac{e_i^2}{\gamma_i-\bm{n}\cdot\bm{u}_i} 
    \left(\frac{d}{d\tau_i}\left[\frac{\bm{n}\times(\bm{n}\times\bm{u}_i)}{\gamma_i-\bm{n}\cdot\bm{u}_i}\right] \right)^2 .
    \label{eq:fsolidangle}
\end{equation}
The first term inside each integral corresponds to the retarded field, and the second term corresponds to the advanced field. Now, we recognize that the integrated flux of the advanced field is identical to that of the retarded field except that it propagates in the opposite direction $f_i\left(\bm{n}, -\bm{u}_i\right) = f_i\left(-\bm{n}, \bm{u}_i\right)$. This property allows one to conclude that $\Delta P^\mu_+=0$ after integrating over all solid angles~\footnote{Note the importance of also taking the integral over all times, which allows for the change of variables below Eq.~\eqref{eq:dPdomega}.}. In Appendix~\ref{ap:solidangle}, we have verified that $\Delta \mathcal{E}_+=0$ by numerically integrating Eq.~\eqref{eq:E+} over virtually all solid angles, by using the trajectories from our simulations in Sec.~\ref{sec:simulations}.

Now, if we consider a distant surface enclosing a neutral system of charges, which are nearly copropagating with each other before and after the collision with a laser pulse, and we integrate the total flux across the surface over time, then the energy-momentum conservation law as in Eq.~\eqref{eq:Ttotalconserv} applies. To simplify further, recall that the total field is defined as the sum of the external (laser) and retarded field $F^{\mu\nu}(x) = F^{\mu\nu}_{\text{ext}}(x) + F^{\mu\nu}_{\text{ret}}(x)$~[see Eq.~\eqref{eq:Atot}]. We expect that the energy contained in the incident laser pulse is approximately the same as that of the outgoing pulse, i.e. that the net transfer of energy between the laser pulse and particles is small. 

In practice, we will approximate the laser pulse as a plane wave in our simulations. By considering the propagation direction of the plane-wave pulse, and the fact that the limit of an infinitely large integration surface will be taken, we expect a negligible contribution from the interference terms of the plane-wave pulse and the retarded field on the integration surface. Therefore, we assume that terms involving the external field are negligibly small compared to those that involve the retarded field, such that Eq.~\eqref{eq:Ttotalconserv} simplifies to
\begin{equation}
    \Delta P^\mu_p \approx -\int_\Sigma T^{\mu\nu}(F_{\text{ret}}, F_{\text{ret}}) d^3\sigma_\nu.
    \label{eq:consv_law_ret}
\end{equation}
This approximation has been verified by demonstrating that the timelike component is satisfied by our simulations, when solving the LL equation with the total field~(see Tab.~\ref{tab:energy_consv}). Therefore, we are left with the familiar statement that the energy and momentum carried away by electromagnetic radiation must correspond to a decrease in the energy and momentum of the particles.

\section{Simulations} 
\label{sec:simulations}

\noindent
In the previous section, we derived a simple generalization of the LL equation, and showed that the energy-momentum conservation law of the HRD action is physically sensible. It is well known that the LL equation satisfies energy-momentum conservation when evaluated with the external field, if the radiation emitted is incoherent. Here we will show that the LL equation satisfies energy-momentum conservation when evaluated with the total field, when the radiation emitted is coherent.

To this end, we simulate the collision of a relativistic $e^-/e^+$ bunch and a laser pulse. As mentioned in the introduction, it is not clear how previous \textit{ad hoc} models of RR apply to an $e^-/e^+$ bunch, which highlights our interest in this particular system. An $e^-/e^+$ bunch is also advantageous because of its inherent stability at the high particle densities needed for coherent emission. In contrast, an $e^-$ bunch would suffer from Coulomb repulsion and high emittance. Note that relativistic $e^-/e^+$ beams are often produced in the laboratory via the Bethe-Heitler process, by ultrarelativistic ($\gamma_0\gg1$) beams of charged particles passing through a high-Z target~\cite{chen2010,sarri2015,arrowsmith2024}. Selection and transport of high quality positron beamlets has recently been shown to be possible~\cite{streeterSR24}, and dense subfemtosecond quasimonoenergetic GeV positron bunches with tens of picocoulombs of charge have also been theoretically demonstrated by colliding a twisted laser pulse with a Gaussian laser pulse~\cite{zhaoCP22}.

\subsection{Numerical code}

\noindent
In our code~\cite{quin2023_phd}, we proceed from first principles by simulating point particles. Alternatively, one might use a particle-in-cell (PIC) code with a high resolution, where many particles are represented by a macroparticle~\cite{vranic2016}. To initialize the code, the $e^-/e^+$ bunch is assumed to propagate ballistically before colliding with the laser pulse (a sensible assumption for a neutral bunch). The interparticle fields are evaluated at the retarded time(s) by interpolating the historical trajectories, which are stored in the memory. Then, either the (reduced) LL~\cite{tamburini2010} or Lorentz~\cite{tamburini2011_phd} equation is integrated by a second-order leapfrog scheme. By this method, the trajectories can be determined at all times, and the spectrum of energy radiated can be evaluated via a fast Fourier transform applied to Eqs.~\eqref{eq:energy_coh} and \eqref{eq:energy_incoh}. 

Each simulation is then characterized by (i) the equation of motion and (ii) the field configuration used. For (i), we can choose whether or not to switch off the self-force by selecting either the (reduced) LL or Lorentz equation. For (ii), these equations can be evaluated with interparticle fields, as a function of $\mathcal{F}^{\mu\nu}(x_i)$, or without interparticle fields, as a function of $F^{\mu\nu}_{\text{ext}}(x_i)$ alone. As our code has the ability to `switch off' the interparticle fields, we can isolate their impact from the external laser field. 

\subsection{Simulation parameters}
\label{sec:sim_params}

\noindent
An ultrarelativistic particle colliding with an intense, linearly polarized laser pulse will emit a quasicontinuous series of harmonics on-axis, starting from the first harmonic ${\lambda_1=\lambda_0\left(1+\frac{1}{2}a_0^2\right)/4\gamma^2_0}$~\cite{sarachik1970, salamin1996}. Here we define the normalized amplitude $a_0=|e|F_0/m\omega_0$, central frequency $\omega_0=2\pi/\lambda_0$ and field amplitude $F_0$ of the laser pulse. To maximize the coherence and impact of the interparticle fields, the radiation emitted should be concentrated at low frequencies ($a_0>1$) while avoiding backscattering of the bunch ($a_0<2\gamma_0$). This suggests that we consider the regime $a_0=\gamma_0$. The first-principles code described above is exact but memory-intensive, and hence we are restricted to simulating a bunch of full-width-at-half-maximum FWHM$_0<\lambda_1$, containing a few thousand particles.

Therefore, we consider a neutral bunch of width FWHM$_0=16\,\si{\nano\metre}$, containing 4000\,$e^-$ and 4000\,$e^+$, and propagating along $+z$ with $\gamma_0=5$. The kinetic energy spread and divergence are $\sigma_{\text{KE}}=0.1\%$ and $\sigma_\vartheta=1\,\si{\milli\radian}$, respectively. This bunch collides head-on with a plane wave pulse
\begin{align}
	\frac{|e|}{m}\bm{A}_{\text{ext}}(\varphi) &= a(\varphi)\,\sin(\varphi)\,\hat{\bm{x}},
	\\
	a(\varphi) &= a_0\cos^2(\varphi/\Delta),
	\label{eq:envelope}
\end{align}
where ${\varphi\in[-\pi\Delta/2, +\pi\Delta/2]}$ is the envelope domain and $\varphi=\omega_0(t+z)$ is the wave phase. A previous simulation carried out with a focused laser pulse, where the waist was $4\,\si{\micro\metre}$, did not show appreciable differences with respect to the plane-wave result.

We consider two lasers with a central wavelength $\lambda_0=[100,~50]\,\si{\nano\metre}$. The normalized amplitude $a_0=5$ and pulse length $\text{FWHM}_L\approx26.7\,\si{\femto\second}$ are the same for both lasers, but the cycle-averaged peak intensity $I_0\approx[3.5,~13.9]\,\times10^{21}\,\si{\watt/\centi\metre^2}$ and parameter $\Delta\approx[440,~880]$ vary for each wavelength respectively. Progress has been made toward lasers operating at $\lambda_0=100\,\si{\nano\metre}$~\cite{drescher2021}, and free electron lasers can produce radiation at $\lambda_0=50\,\si{\nano\metre}$ by tuning the electron energy accordingly~\cite{bostedt2016,huang2007}. Finally, note that a small time step $\Delta t\approx 0.27\,\text{as}$ was needed to resolve the radiation spectrum.

As will become clear, these simulations are dominated by the external laser field as opposed to the interparticle fields. It therefore makes little difference whether we evaluate the quantum parameter $\chi_i(x_i)$ with the total or external field. Considering the amplitude of the laser alone, we conclude that QED effects are negligible here $\chi_0=2\gamma_0F_0/F_{\text{cr}}\approx[1.2,~2.4]\times10^{-3}$. Note that the average interparticle distance in the rest frame $d\approx24\,a_{\text{ps}}$ is well above the Bohr radius for positronium $a_{\text{ps}}\approx0.11$\,\si{\nano\meter}, indicating that bound state formation and annihilation are unlikely~\cite{quin2023_phd}. In a plane wave, the self-force of the LL equation is proportional to $R_C=\alpha \chi_0 a_0$~\cite{dipiazza2012_review}, where $\alpha\approx1/137$ is the fine structure constant. Then, $R_C\gtrsim1$ and $\chi_0\ll1$ is referred to as the classical radiation dominated regime, though we are far from this regime here $R_C\approx[4.4,~8.8]\times10^{-5}$.

\begin{figure}[t]
    \centering
    \includegraphics[width=0.999\linewidth]{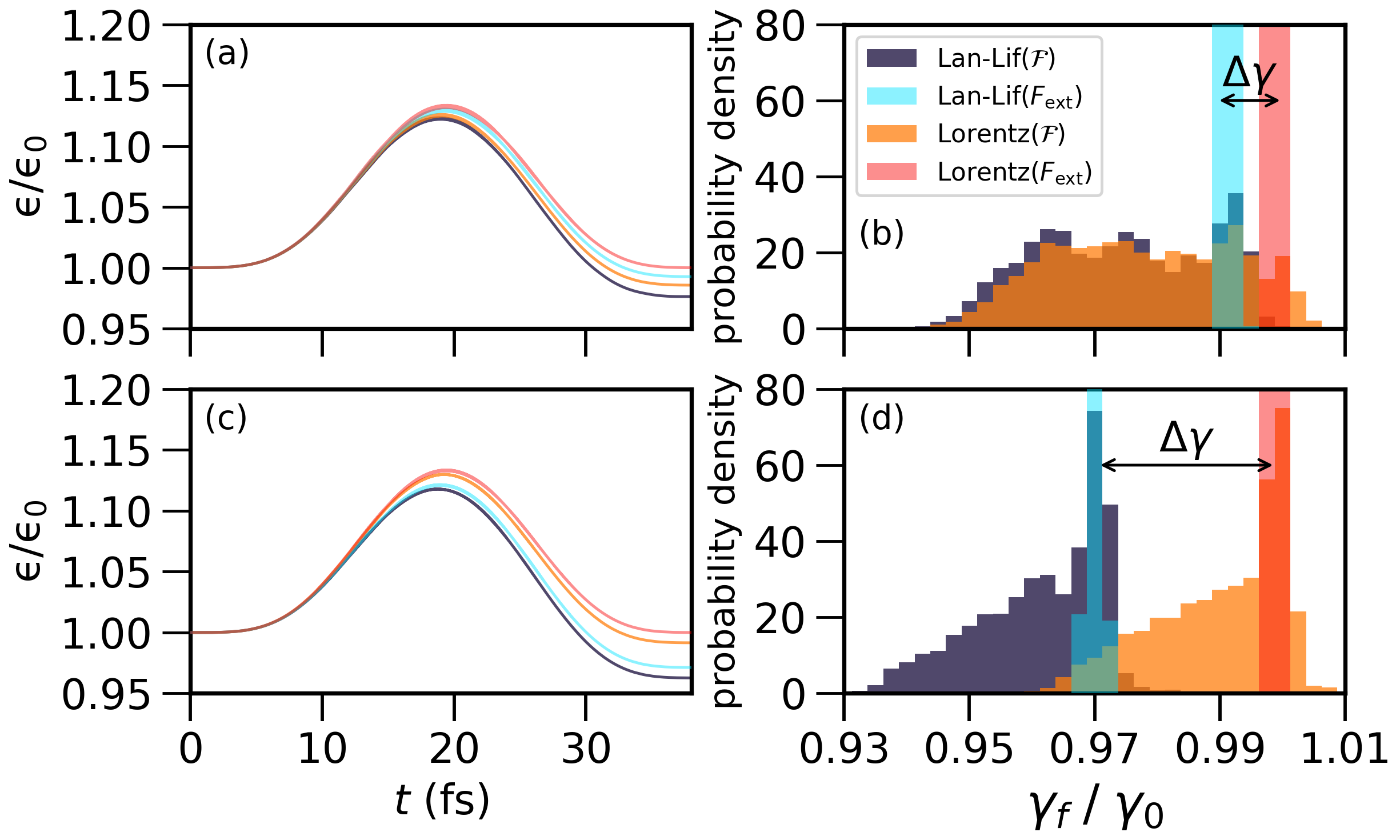}
    \caption{Total particle energy over time (a,\,c) and probability density of final Lorentz factors (b,\,d) in the collision of an $e^-/e^+$ bunch and laser pulse. Wavelength is (a,\,b) $\lambda_0=100\,\si{\nano\metre}$, (c,\,d) $\lambda_0=50\,\si{\nano\metre}$. Here the Lorentz and reduced LL (Lan-Lif) equations are evaluated with either the total or external field. $\Delta\gamma$ is the change in Lorentz factor due to the external field predicted by  Eq.~\eqref{eq:dgamma}. Legend in (b) applies to all plots. Rapid oscillations due to linear polarization have been removed via a moving average in (a,\,c). Area under (b,\,d) is normalized to unity.
    }
    \label{fig:energy_loss}
\end{figure}

\subsection{Inhomogeneous energy loss}

\noindent
The evolution of the total energy $\epsilon(t)=\sum^N_{i=1}m_i\gamma_i(t)$ during the collision of the $e^-/e^+$ bunch with the laser pulse is shown in Fig.~\ref{fig:energy_loss}\,(a,\,c). Here, one can see that solving the LL or Lorentz equation with either the total or external field will lead to a different energy loss in each case. The magnitude of the energy loss $\Delta\epsilon=|\epsilon_f-\epsilon_0|$ is shown in Tab.~\ref{tab:energy_consv}, where $\epsilon_0\equiv\epsilon(t_0)$ and $\epsilon_f\equiv\epsilon(t_f)$ are the initial and final energy respectively, and $\epsilon_0\approx20.4\,\si{\giga\electronvolt}$ in all of our simulations. 

In general, the energy loss tends to increase when both the self-force and interparticle fields are included~[Fig.~\ref{fig:energy_loss}\,(a,\,c)]. However, the energy loss caused by the interparticle fields is inhomogeneous~[Fig.~\ref{fig:energy_loss}\,(b,\,d)]. This inhomogeneity appears to result from the anisotropic emission of radiation by relativistic particles. Radiation emitted from particles near the rear of the bunch will only be felt by particles at the front of the bunch. As each particle experiences a different field, it will lose a different amount of energy and follow a different trajectory.

An inhomogeneous energy loss indicates an expansion in phase space and increase in entropy. Previous works in the literature on classical RR predict a contraction of phase space and decrease in entropy, when solving the \textit{collisionless} Vlasov equation coupled to the (classical) LL equation~\cite{hazeltine2004, tamburini2011, burton2014}. There are a few possible reasons for this discrepancy between our results and the prior literature. Usually one does not consider interparticle fields when solving the Vlasov equation, but rather a mean field found by sampling the charge and current distribution and solving Maxwell's equations~(as in a PIC code). In addition, collisional terms might become important for the range of parameters considered here. Note that the stochastic nature of photon emission in QED will give rise to an increase in entropy~\cite{neitz2013}, though this effect is not relevant here ($\chi_0\ll 1$).

To understand the effect of the interparticle fields on the energy loss, we should first isolate the impact of the external field. For this, we note that the LL can be solved exactly in a plane wave~\cite{dipiazza2008}. From this analytical solution, we can predict the change in Lorentz factor $\Delta\gamma=|\gamma_f-\gamma_0|$ due to the external field alone, for an approximately ultrarelativistic particle colliding head-on with the plane wave pulse in Eq.~\eqref{eq:envelope}, as
\begin{equation}
    \frac{\Delta\gamma}{\gamma_0} \approx \frac{\frac{1}{8}\pi R_C\Delta}{1+\frac{1}{8}\pi R_C\Delta}.
    \label{eq:dgamma}
\end{equation}
This estimate $\Delta\gamma/\gamma_0=[0.008,~0.029]$ has been plotted in Fig.~\ref{fig:energy_loss}\,(b,\,d). Here, when solving the LL equation with the external field, each particle has the same change in Lorentz factor equal to $\Delta\gamma$. We also notice that the energy losses caused by solving the LL equation with the total field are nearly the same as those observed when solving the Lorentz equation with the total field, except that the particles' Lorentz factors are decreased by $\Delta\gamma$. This suggests that the self-force is approximately the same for each particle, because the external field dominates over the interparticle fields. We conclude that the energy loss induced by the interparticle fields must occur primarily through the Lorentz force, for the simulation parameters studied here.

\subsection{Radiation spectrum}

\label{sec:spectrum}

\begin{figure}[t]
    \centering
    \includegraphics[width=0.999\linewidth]{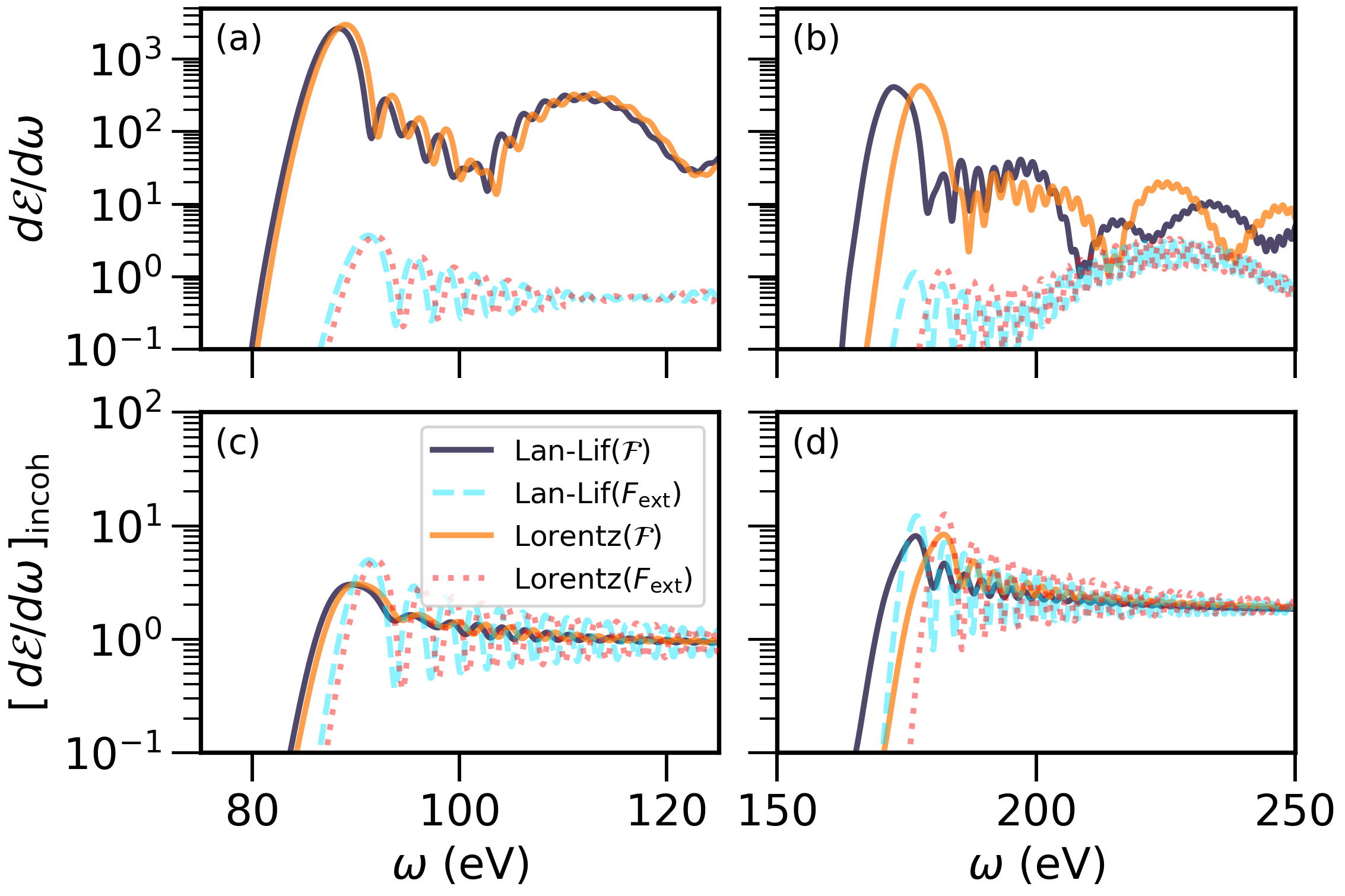}
    \caption{Total (a,\,b) and incoherent (c,\,d) spectrum of energy radiated onto a small detector from an $e^-/e^+$ bunch colliding with a laser pulse. Wavelength is (a,\,c) $\lambda_0=100\,\si{\nano\metre}$ and (b,\,d) $\lambda_0=50\,\si{\nano\metre}$. Rapid oscillations have been removed by a moving average of $2\,\si{\electronvolt}$. Legend in (c) applies to all plots.}
    \label{fig:spec_detector}
\end{figure}

\noindent
Having determined the impact of the interparticle fields on the energy lost, we turn our attention to the radiation spectrum. This can be constructed from the Fourier transform of the retarded fields~(see Appendix~\ref{ap:threevector}, and Ref.~\cite{jackson1998})
\begin{equation}
     \frac{d\mathcal{E}}{d\omega d\Omega} = \frac{1}{4\pi^2} \left|\sum^N_{i=1}\int^{+\infty}_{-\infty} \frac{d\mathcal{A}^\mu_i}{d\tau_i} e^{i\omega(nx_i)} d\tau_i \right|^2,
    \label{eq:energy_coh}
\end{equation}
for a distant observer in the direction of unit-vector $\bm{n}$, where $n^\mu\equiv n^\mu_+ = (1,\bm{n})$ is a null four-vector satisfying $(n)^2=0$. The function ${\mathcal{A}^\mu_i(\tau_i) = e_iu^\mu_i/(nu_i)}$ is essentially the retarded potential multiplied by $\mathcal{R}$. We can also define the incoherent part of the spectrum as
\begin{equation}
    \frac{d\mathcal{E}}{d\omega d\Omega}\Bigg|_{\text{incoh}} = \frac{1}{4\pi^2}\sum^N_{i=1}\Bigg| \int^{+\infty}_{-\infty} \frac{d\mathcal{A}^\mu_i}{d\tau_i} e^{i\omega(nx_i)} d\tau_i \Bigg|^2.
    \label{eq:energy_incoh}
\end{equation}
In our code, these integrals are solved numerically via a fast Fourier transform with the trajectories from our simulations. The resulting spectra are plotted in Fig.~\ref{fig:spec_detector}, and have been integrated in the solid angle over a 1\,\si{\centi\metre^2} detector placed at 1\,\si{\metre} along the $+z$ axis. Here, we have plotted a range of frequencies around the first harmonic $\omega_1\approx[90,~180]\,\si{\electronvolt}$ for each wavelength $\lambda_0=[100,~50]\,\si{\nano\metre}$ respectively. Note that the first harmonic tends to scale coherently and dominate the system.

First, we consider the spectrum emitted by the $e^-/e^+$ bunch colliding with the $\lambda_0=100\,\si{\nano\metre}$ laser, in Fig.~\ref{fig:spec_detector}\,(a,\,c). The choice of the equation of motion, whether Lan-Lif or Lorentz, has a relatively small impact here, and thus the effect of the
self-force is small (on a log scale). Yet the field configuration does have a significant impact. By solving either equation of motion with the total field, the energy radiated at the first harmonic increases by about two orders of magnitude. The energy loss induced by the interparticle fields via the Lorentz force also redshifts the spectrum to lower frequencies. By comparing the total~[Fig.~\ref{fig:spec_detector}\,(a)] and incoherent~[Fig.~\ref{fig:spec_detector}\,(c)] spectra, it is clear that this additional energy loss occurs due to increased coherence as opposed to a change in the underlying trajectories.

Now consider the spectrum emitted when the bunch collides with the $\lambda_0=50\,\si{\nano\metre}$ laser, in Fig.~\ref{fig:spec_detector}\,(b,\,d). Here we can see the self-force redshifts the spectrum by about 5\,\si{\electronvolt}, but does not appear to significantly alter the amplitude of the spectrum (on a log scale). This effect occurs in both the total~[Fig.~\ref{fig:spec_detector}\,(b)] and incoherent spectra~[Fig.~\ref{fig:spec_detector}\,(d)], which indicates a change in the underlying trajectories of the particles. Once again, we see that the interparticle fields induce an increase in the energy radiated via the Lorentz force, due to coherent emission.

\subsection{Energy conservation}

{
\setlength{\tabcolsep}{1.2em} % horizontal padding
\renewcommand{\arraystretch}{1.4} % vertical padding
\begin{table}[t]
\centering
\begin{tabular}{@{}lccc@{}}
\toprule
& $\Delta\epsilon/\epsilon_0$ & $\Delta\mathcal{E}/\epsilon_0$ & $[\Delta\mathcal{E}]_{\text{incoh}}/\epsilon_0~$\\ 
\midrule
$\lambda_0=100\,\si{\nano\metre}$\\
Lan-Lif($\mathcal{F}$) & 0.024 & 0.025 & 0.007\\
Lan-Lif($F_{\text{ext}}$) & 0.007 & 0.013 & 0.007\\
Lorentz($\mathcal{F}$) & 0.014 & 0.024 & 0.007 \\
Lorentz($F_{\text{ext}}$) & 0.000 & 0.012 & 0.007\\
\midrule
$\lambda_0=50\,\si{\nano\metre}$\\
Lan-Lif($\mathcal{F}$) & 0.037 & 0.035 & 0.027\\
Lan-Lif($F_{\text{ext}}$) & 0.029 & 0.027 & 0.027\\
Lorentz($\mathcal{F}$) & 0.008 & 0.035 & 0.027\\
Lorentz($F_{\text{ext}}$) & 0.000 & 0.027 & 0.027\\
\bottomrule 
\end{tabular}
\caption{\label{tab:energy_consv} 
Energy lost by particles $\Delta\epsilon$, total $\Delta\mathcal{E}$ and incoherent $[\Delta\mathcal{E}]_{\text{incoh}}$ energy radiated from the collision of an $e^-/e^+$ bunch and laser pulse. Initial energy $\epsilon_0$ of the bunch is constant.  We solve either the reduced LL (Lan-Lif) or Lorentz equation as a function of the total or external field to determine the energy lost. These trajectories are used to calculate the energy radiated over all angles (see Appendix~\ref{ap:solidangle}).
}
\end{table}
}

\noindent
Earlier, we obtained a conservation law which relates the change in energy and momentum of the particles, to the energy and momentum carried away by the retarded fields~[see Eq.~\eqref{eq:consv_law_ret}]. Although this result was derived by integrating over all times, in practice we only need to integrate over the finite time interval during which the particles' acceleration is nonzero. Here we will check numerically whether the timelike component, that is the energy conservation law, is satisfied for each of our simulations. For a simulation to be consistent, we require that the energy lost by the particles is equal to the energy radiated. 

The energy $\Delta\epsilon$ lost by the particles can be directly obtained from the simulation results shown in Fig.~\ref{fig:energy_loss}. To obtain the spectrum of energy radiated, we can evaluate Eqs.~\eqref{eq:energy_coh} and \eqref{eq:energy_incoh} as described above; except now we must integrate over virtually all angles, as opposed to a small range of angles around the $z$-axis. This spectrum is shown in Appendix~\ref{ap:solidangle}, and can be integrated to estimate the total $\Delta\mathcal{E}$ and incoherent $[\Delta\mathcal{E}]_{\text{incoh}}$ energy radiated. 

Therefore, one can find estimates in Tab.~\ref{tab:energy_consv} for the energy lost and energy radiated in each simulation. In general, the best agreement between the energy lost and total energy radiated occurs when solving the LL equation with the total field, as implied by the HRD action. Note that this verifies the assumptions we made when deriving Eq.~\eqref{eq:consv_law_ret}, i.e., that we can ignore any net exchange of energy between the external field and the particles. Any remaining discrepancy between $\Delta\epsilon$ and $\Delta\mathcal{E}$ can likely be attributed to the numerical resolution from the integration over the solid angle, and from the truncation of the spectrum at high frequencies above the Nyquist value.

When solving the LL equation with the external field in Tab.~\ref{tab:energy_consv}, we note that there is good agreement between the energy lost and \textit{incoherent} energy radiated. The energy lost in this case is roughly equal to $\Delta\gamma/\gamma_0=[0.008,~0.029]$ for each wavelength as predicted by Eq.~\eqref{eq:dgamma}. In other words, evaluating the LL equation with the external field properly accounts for incoherent emission, but does not account for coherent emission. Once again, we conclude that the LL equation must be evaluated with the total field, to properly account for the work done by the interparticle fields during coherent emission.

Solving the Lorentz equation with the total field provides a reasonable estimate of the total energy radiated in Tab.~\ref{tab:energy_consv}, but does not properly account for the energy lost by the particles. This method includes the work done by the interparticle fields via the Lorentz force, but neglects the energy loss caused by the self-force. Meanwhile, evaluating the Lorentz force with the external field fails to account for any energy loss whatsoever, violating energy-momentum conservation.

In summary, the interparticle fields induce an energy loss in our simulations via the Lorentz force~[Fig.~\ref{fig:energy_loss}]. This is in addition to the energy loss caused by the self-force, which is effectively dominated by the external field, and roughly the same for all particles. From the radiation spectrum~[Fig.~\ref{fig:spec_detector}], we can see that the additional energy loss caused by the interparticle fields occurs due to increased coherence. Here, the energy loss induced by the self-force and interparticle fields can redshift the radiation spectrum. Finally, we have shown in Tab.~\ref{tab:energy_consv} that solving the LL equation with the total field is a consistent approach for the range of parameters studied here, where the external field dominates over the interparticle fields.

\section{Conclusion}

\noindent
In this article, we reviewed the equations of motion and conservation laws which result from the Hamilton-Rohrlich-Dirac (HRD) action. A simple generalization of the LL equation was presented, which includes coherence effects via the interparticle fields. A simplified energy-momentum conservation law was derived from the HRD theory, which has a clear interpretation: that the energy lost by the particles and the electromagnetic field must correspond to energy carried away by electromagnetic radiation. Finally, we numerically solved the reduced LL equation and radiation spectra for the collision of an $e^-/e^+$ bunch with a plane wave pulse. In particular, we found that solving the LL equation with the total field provides the best agreement between the energy lost by the particles and energy radiated, and is therefore the most consistent approach (note that in the initially neutral system considered here the energy stored by the electromagnetic field is negligibly small).

Radiation reaction becomes significant for ultrarelativistic particles in the presence of a strong electromagnetic field. In this regime, particles tend to emit incoherently. Therefore, it is challenging to identify regimes where both coherence effects and radiation reaction are important. Notably, this implies a high density and large volume of relativistic particles emitting coherently. Hence, we have proceeded from first principles by studying an $e^-/e^+$ bunch in detail which allowed us to verify the HRD model. In future, it would be interesting to generalize these results to a macroscopically large beam of particles in the presence of an optical laser, but this would likely require a different numerical approach.

\appendix

\section{The Lorentz-Abraham-Dirac Equation}
\label{ap:LAD}

\noindent
Here we outline the key steps needed to derive LAD equation from the HRD action. By carrying out the derivatives in the Euler-Lagrange equation~\eqref{eq:EulerLag}, we obtain
\begin{equation}
    m_i\mathrm{a}^\mu_i = e_i F^{\mu\nu}_f(x_i) u_{i,\nu} + e_i \sum^N_{\substack{j=1\\j\neq i}} F^{\mu\nu}_{+j}(x_i) u_{i,\nu}.
    \label{eq:EoM_Fplus}
\end{equation}
Now, the free field is simply the sum of the external and minus field~[see Eq.~\eqref{eq:Afree}]. Therefore, this equation depends on the plus and minus fields, which is not manifestly causal, and we want to remove this dependence wherever possible. This can be achieved by combining the plus and minus fields from all particles except $i$ to create the retarded field~[see Eq.~\eqref{eq:A_plusminus}]. Therefore, we can write
\begin{equation}
    F^{\mu\nu}_f(x_i) + \sum^N_{\substack{j=1\\j\neq i}} F^{\mu\nu}_{+j}(x_i) = \mathcal{F}^{\mu\nu}_i(x_i) + F^{\mu\nu}_{-\,i}(x_i),
\end{equation}
where $\mathcal{F}^{\mu\nu}_i(x_i)$ is defined in Eq.~\eqref{eq:Ftot} and contains the external and retarded fields. Finally, the LAD self-force is 
\begin{equation}
    e_i F^{\mu\nu}_{-\,i}(x_i) u_{i,\nu} = \frac{2}{3} e_i^2 \left( \dot{\mathrm{a}}^\mu_i + \mathrm{a}^2_i u^\mu_i \right),
    \label{eq:LADselfforce}
\end{equation}
where we have used the expression for $F^{\mu\nu}_{-\,i}(x_i)$ in Eq.~\eqref{eq:F_minus}, the on shell-condition $u^2_i=1$ and its second derivative $(\dot{\mathrm{a}}_iu_i)=-\mathrm{a}^2_i$. Following these steps, we arrive at the LAD equation in Eq.~\eqref{eq:LADeq}. Note that the LAD equation is a well known result which has been derived elsewhere~\cite{dirac1938, abraham1905, teitelboim1970, hamilton1971, rohrlich2007, rohrlich1964, barut1980}. The key step omitted here is the evaluation of $F^{\mu\nu}_{-\,i}(x_i)$, which can be found in textbooks by Rohrlich~\cite{rohrlich2007} and Barut~\cite{barut1980}.

\section{Energy-momentum tensor in the asymptotic limit}
\label{ap:Tasymptotic}

\noindent
To evaluate the integral in Eq.~\eqref{eq:dP+dOmega}, we require expressions for the energy-momentum tensor as a function of the plus field, and hence the retarded and advanced fields, as derived here.

\subsection{Products of retarded and advanced fields}

\noindent
First, consider the cross terms in the energy-momentum tensor, which include products of the advanced and retarded fields
\begin{align}
    T^{\mu\nu}\Big(& F_{\substack{\text{ret}\,i\\\text{adv}\,i}}, F_{\substack{\text{adv}\,i\\\text{ret}\,i}} \Big) \tilde{n}_\nu = \nonumber
    \\[0.5em]
    & \frac{1}{4\pi} \left[F^{\mu\alpha}_{\substack{\text{ret}\,i\\\text{adv}\,i}}\eta_{\alpha\beta}F^{\beta\nu}_{\substack{\text{adv}\,i\\\text{ret}\,i}} \tilde{n}_\nu + \frac{1}{4} \left( F^{\alpha\beta}_{\text{ret}\,i} F^{\phantom{\alpha}}_{\text{adv}\,i,\alpha\beta} \right) \tilde{n}^\mu \right].
    \label{eq:Tcross}
\end{align}
Given the fields in far region (see Eq.~\eqref{eq:Fretadv_asymptotic}), we can write the first term as
\begin{align}
    F^{\mu\alpha}_{\text{ret}\,i} \eta_{\alpha\beta} &F^{\beta\nu}_{\text{adv}\,i} \tilde{n}_\nu = -\Bigg[\Bigg. \frac{4e^2_i}{\mathcal{R}^2(n_+u_i)(n_-u'_i)} \nonumber
    \\[0.5em]
    &\times\frac{\partial^2}{\partial\tau_i\partial\tau'_i} \left( 
    \frac{n^{[\mu}_+ u_i^{\alpha]} \eta_{\alpha\beta} n^{[\beta}_- u'^{\nu]}_i \tilde{n}_\nu}{(n_+u_i)(n_-u'_i)}\right) \Bigg.\Bigg]_{\substack{\tau_i=\tau_{\text{ret}\,i},\\ \tau'_i=\tau_{\text{adv}\,i}}}.
\end{align}
Here the four-velocity $u'^\mu_i\equiv u_i(\tau'_i)$ is defined as a function of the dummy variable $\tau'_i$, and the normal can be written as $\tilde{n}^\mu=\frac{1}{2}(n^\mu_+ + n^\mu_-)$. The expression under the derivative can be written as 
\begin{align}
    \Rightarrow &\frac{(n_+u'_i)u^\mu_i - (u_iu'_i)n^\mu_+}{8(n_+u_i)(n_-u'_i)}(n_-)^2 + \frac{(n_-u_i)n^\mu_+-(n_+n_-)u^\mu_i}{8(n_+u_i)} \nonumber
    \\
    &\hspace{3em} + \frac{(n_-u_i)(n_+u'_i)-(n_+n_-)(u_iu'_i)}{8(n_+u_i)(n_-u'_i)}n^\mu_+.
\end{align}
Note that there are three terms on the right hand side. The first vanishes due to the null property $(n_+)^2=0$, and the second vanishes when differentiated with respect to $\tau'_i$, hence we are left with the third.

The derivation of all other terms in the energy-momentum tensor proceeds by the same manner. We can summarize the results as
\begin{align}
    F^{\mu\alpha}_{\substack{\text{ret}\,i\\\text{adv}\,i}} \eta_{\alpha\beta} F^{\beta\nu}_{\substack{\text{adv}\,i\\\text{ret}\,i}}\tilde{n}_\nu &= -\frac{1}{2}g_i(\tau_{\text{ret}\,i} , \tau_{\text{adv}\,i}) n^\mu_\pm,
    \label{eq:Tcross_eval1}
    \\
    \frac{1}{2} \left( F^{\alpha\beta}_{\text{ret}\,i} F^{\phantom{\alpha\beta}}_{\text{adv}\,i,\alpha\beta} \right) \tilde{n}^\mu &= g_i(\tau_{\text{ret}\,i}, \tau_{\text{adv}\,i}) \tilde{n}^\mu,
    \label{eq:Tcross_eval2}
\end{align}
where we have defined the scalar function
\begin{align}
    g_i(\tau_i, \tau'_i) = &\frac{e^2_i}{\mathcal{R}^2(n_+u_i)(n_-u'_i)} \nonumber 
    \\
    &\hspace{1em} \times \frac{\partial^2}{\partial\tau_i\partial\tau'_i} \left[  \frac{(n_-u_i)(n_+u'_i)+2(u_iu'_i)}{(n_+u_i)(n_-u'_i)}\right],
\end{align}
and used the property $(n_-n_+)=-2$. Finally, we recognize that the sum of Eq.~\eqref{eq:Tcross_eval2} and both components of Eq.~\eqref{eq:Tcross_eval1} are zero. Consequently, the cross terms in the energy-momentum tensor are identically zero~[see Eq.~\eqref{eq:Tcross_zero}]. 

\subsection{Square of retarded and advanced fields}

\noindent
Now we can compute the energy-momentum tensor as a function of the square of the retarded and advanced fields
\begin{align}
    T^{\mu\nu}\Big(& F_{\substack{\text{ret}\,i\\\text{adv}\,i}}, F_{\substack{\text{ret}\,i\\\text{adv}\,i}} \Big) \tilde{n}_\nu = \nonumber
    \\[0.5em]
    & \frac{1}{4\pi} \left[F^{\mu\alpha}_{\substack{\text{ret}\,i\\\text{adv}\,i}}\eta_{\alpha\beta}F^{\beta\nu}_{\substack{\text{ret}\,i\\\text{adv}\,i}} \tilde{n}_\nu + \frac{1}{4} \left( F^{\alpha\beta}_{\substack{\text{ret}\,i\\\text{adv}\,i}} F^{\phantom{\alpha}}_{\substack{\text{ret}\,i\\\text{adv}\,i}\,,\,\alpha\beta} \right) \tilde{n}^\mu \right].
    \label{eq:Tsquare}
\end{align}
As before, we use the fields in the asymptotic limit to construct the first term
\begin{align}
    F^{\mu\alpha}_{\text{ret}\,i} \eta_{\alpha\beta} &F^{\beta\nu}_{\text{ret}\,i} \tilde{n}_\nu = \Bigg[\Bigg. \frac{4e^2_i}{\mathcal{R}^2(n_+u_i)(n_+u'_i)} \nonumber
    \\[0.5em]
    &\times\frac{\partial^2}{\partial\tau_i\partial\tau'_i} \left( 
    \frac{n^{[\mu}_+ u_i^{\alpha]} \eta_{\alpha\beta} n^{[\beta}_+ u'^{\nu]}_i \tilde{n}_\nu}{(n_+u_i)(n_+u'_i)}\right) \Bigg.\Bigg]_{\substack{\tau_i=\tau_{\text{ret}\,i},\\ \tau'_i=\tau_{\text{ret}\,i}}}.
\end{align}
The term under the derivative can be written as
\begin{align}
    \Rightarrow& - \frac{(n_+n_-)(u_iu'_i)n^\mu_+}{8(n_+u_i)(n_+u'_i)} - \frac{(u_iu'_i)n^\mu_+ + (n_-u'_i)u^\mu_i}{8(n_+u_i)(n_+u'_i)}(n_+)^2 \nonumber
    \\
    &\hspace{2em} + \frac{n^\mu_+}{8} + \frac{(n_-u'_i)n^\mu_+}{8(n_+u'_i)} + \frac{(n_+n_-)u^\mu_i}{8(n_+u_i)}.
\end{align}
We recognize five terms on the right hand side. The first can be simplified using $(n_-n_+)=-2$. The second vanishes due to the null property $(n_-)^2=0$. The third, fourth and fifth terms are constants or functions of one variable (either $\tau_i$ or $\tau'_i$), and therefore vanish under the second derivative. Hence, only the first term remains.

The derivation proceeds in the same manner for the square of the advanced field, and we can summarize the results as
\begin{align}
    F^{\mu\alpha}_{\substack{\text{ret}\,i\\\text{adv}\,i}} & \eta_{\alpha\beta} F^{\beta\nu}_{\substack{\text{ret}\,i\\\text{adv}\,i}}\tilde{n}_\nu = \rho_{\pm i}\left(\tau_{\substack{\text{ret}\,i\\ \text{adv}\,i}}\right) n^\mu_\pm,
    \\
    \rho_{\pm i}(\tau_i) &= \left(\frac{e_i}{\mathcal{R}(n_\pm u_i)} \frac{d}{d\tau_i} \left[\frac{u^\mu_i}{(n_\pm u_i)} \right] \right)^2 . 
\end{align}
By a similar method, one can show using the null property $(n_\pm)^2=0$ that the second term of the energy-momentum tensor vanishes
\begin{equation}
    F^{\alpha\beta}_{\text{ret}\,i} F^{\phantom{\alpha}}_{\text{ret}\,i,\alpha\beta} = F^{\alpha\beta}_{\text{adv}\,i} F^{\phantom{\alpha}}_{\text{adv}\,i,\alpha\beta} = 0,
\end{equation}
and so we arrive at the expression for the energy-momentum tensor in Eq.~\eqref{eq:Tret2}.

\section{Changing from covariant to three-vector notation}
\label{ap:threevector}

\noindent
Here we demonstrate how one can change from covariant to three-vector notation, either in our expressions for the radiation spectra
or the energy-momentum tensor. First, consider the radiation spectrum in Eq.~\eqref{eq:energy_coh}. If we expand the square modulus we will obtain a double sum, and a double integral, over terms like
\begin{equation}
    \frac{d\mathcal{A}^\mu_i}{d\tau_i} \frac{d\mathcal{A}_{j,\mu}}{d\tau_j} = e_ie_j\frac{\partial^2}{\partial\tau_i\partial\tau_j} \left[ \frac{(u_iu_j)}{(nu_i)(nu_j)} \right].
\end{equation}
where $u_j\equiv u_j(\tau_j)$ and $n^\mu=n^\mu_+$. Using the following identity
\begin{align}
    \frac{(u_iu_j)}{(n_\pm u_i)(n_\pm u_j)} = &\frac{(\bm{n}\cdot\bm{u}_i)(\bm{n}\cdot\bm{u}_j)- \bm{u}_i\cdot\bm{u}_j}{(n_\pm u_i)(n_\pm u_j)} \nonumber
    \\
    &\hspace{1em} \pm \frac{\gamma_i}{(n_\pm u_i)} \pm \frac{\gamma_j}{(n_\pm u_j)} - 1,
\end{align}
we recognize the second, third and fourth terms on the right hand side are constants, or a function of one variable (either $\tau_i$ or $\tau_j$), and will therefore vanish under the second derivative. The first term can be simplified using
\begin{align}
     (\bm{n}\cdot\bm{u}_i)&(\bm{n}\cdot\bm{u}_j) - \bm{u}_i\cdot\bm{u}_j = \nonumber
     \\
     &-[\bm{n}\times(\bm{n}\times\bm{u}_i)] \cdot [\bm{n}\times(\bm{n}\times\bm{u}_j)].
\end{align}
Combining these identities, we recognize that the radiation spectrum~\eqref{eq:energy_coh} can be written in a more familiar form~\cite[Eq.\,(14.65)]{jackson1998}
\begin{equation}
    \frac{d\mathcal{E}}{d\omega d\Omega} = \frac{1}{4\pi^2} \left|\sum^N_{i=1}\int^{+\infty}_{-\infty} \frac{d}{d\tau_i}\left[\frac{\bm{n}\times(\bm{n}\times\bm{u}_i)}{\gamma_i-\bm{n}\cdot\bm{u}_i}\right] e^{i\omega(nx_i)} d\tau_i \right|^2.
\end{equation}
Similarly, we can write Eq.~\eqref{eq:hpm} in three-vector notation
\begin{equation}
    \rho_{\pm i}(\tau_i) = \left(\frac{e_i}{\mathcal{R}(\gamma_i\mp \bm{n}\cdot\bm{u}_i)} \frac{d}{d\tau_i} \left[\frac{\bm{n}\times(\bm{n}\times\bm{u}_i)}{\gamma_i \mp \bm{n}\cdot\bm{u}_i} \right] \right)^2 .
\end{equation}
From this result, one can obtain $f_i(\bm{n}, \bm{u}_i)$ as written in Eq.~\eqref{eq:fsolidangle}.

\section{Numerical integration over the solid angle}
\label{ap:solidangle}

\subsection{Retarded fields} 

\begin{figure}[t]
    \centering
    \includegraphics[width=0.99\linewidth]{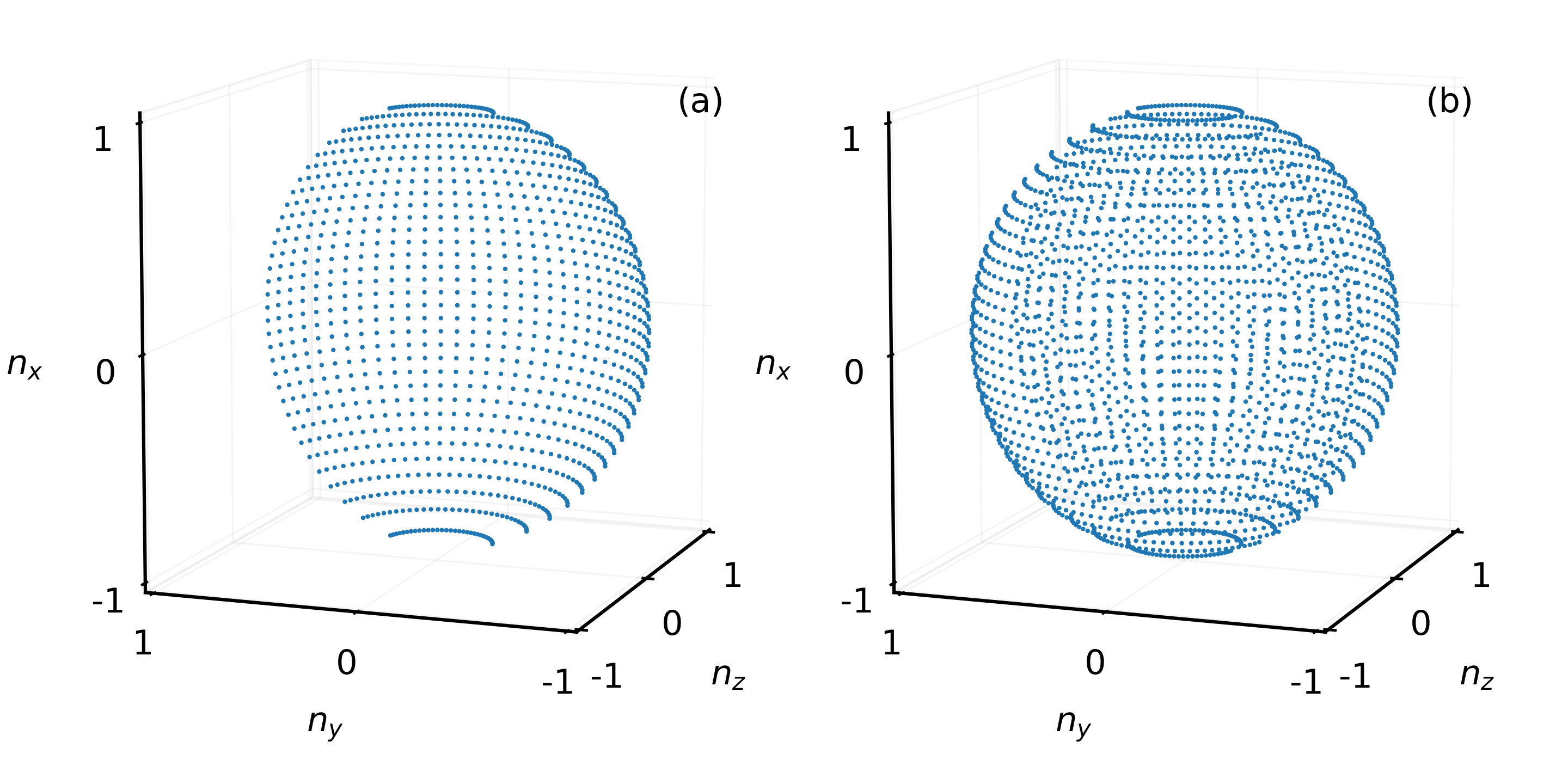}
    \caption{(a) Hemispherelike surface over which the flux of the retarded fields is integrated in the solid angle to obtain Fig.~\ref{fig:spec_detector}. (b) Spherelike surface over which the flux of the retarded and advanced fields is integrated, to demonstrate that Eq.~\eqref{eq:E+} is zero. Observation direction $\bm{n}=(n_x, n_y, n_z)$ satisfies $|\bm{n}|=1$.
    }
    \label{fig:detector}
\end{figure}

\noindent
To demonstrate the consistency of our simulation results, we need to integrate the radiation spectra in Eqs.~\eqref{eq:energy_coh} and~\eqref{eq:energy_incoh} over virtually all solid angles. In our simulations, we expect that most of the radiation will be emitted in a cone of half-angle $|u_x/u_z|\sim a_0/\gamma_0=1\,\si{\radian}$ around the $+z$ axis~(see trajectory in a plane wave in Ref.~\cite[\S\,47]{landaulifshitz_vol2}). In practice, a slightly larger half-angle $\vartheta=1.3\,\si{\radian}$ was needed for numerical convergence (note that this is almost a hemisphere). Therefore, the radiation spectrum was integrated over a patch on the unit sphere of domain $\vartheta_{xz}, \vartheta_{yz}\in[-\vartheta, +\vartheta]$, where $\vartheta_{xz}$ and $\vartheta_{yz}$ are angles measured from the $+z$ axis in the $xz$ and $yz$ planes respectively. In total, 1089 points were distributed uniformly on this surface, as seen in Fig.~\ref{fig:detector}\,(a). 

\begin{figure}[t]
    \centering
    \includegraphics[width=0.999\linewidth]{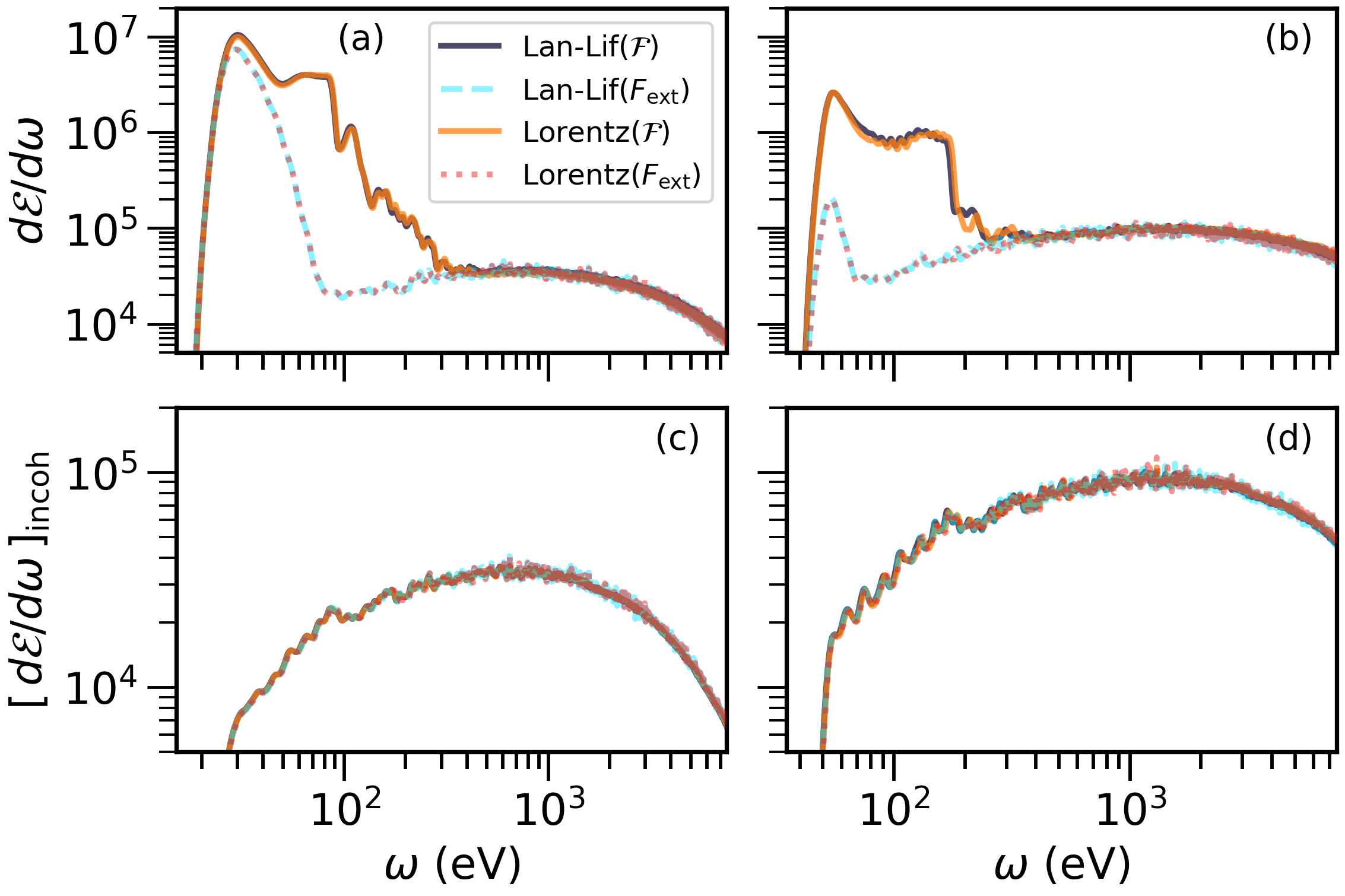}
    \caption{Total (a,\,b) and incoherent (c,\,d) spectrum of energy radiated over all solid angles, from an $e^-/e^+$ bunch colliding with a laser pulse. Wavelength is (a,\,c) $\lambda_0=100\,\si{\nano\metre}$ and (b,\,d) $\lambda_0=50\,\si{\nano\metre}$. Rapid oscillations have been removed by a moving average of $5\,\si{\electronvolt}$. Spectra correspond to energy loss in Fig.~\ref{fig:energy_loss}, legend in (a) applies to all plots.}
    \label{fig:spec_allsolidangles}
\end{figure}

After integrating over this surface we obtain the energy radiated per unit frequency, plotted in Fig.~\ref{fig:spec_allsolidangles} up to the Nyquist value $\pi/\Delta t\approx 7700\,\si{\electronvolt}$. At first glance, there is no visible dependence on the equation of motion (Lorentz or LL). This is because we have integrated over virtually all solid angles and presented nearly the entire range of frequencies on a logarithmic scale. The effect of the self-force can be more clearly seen in Fig.~\ref{fig:spec_detector}, where we plot a small range of frequencies around the first harmonic, and integrate over a small $1\,\si{\centi\metre^2}$ detector. By numerically integrating Fig.~\ref{fig:spec_allsolidangles}, we can estimate the total $\Delta\mathcal{E}$ and incoherent $[\Delta\mathcal{E}]_{\text{incoh}}$ energy radiated, as shown in Tab.~\ref{tab:energy_consv}.

\subsection{Retarded and advanced fields} 

\noindent
Now, consider the energy of the plus field per unit solid angle as shown in Eq.~\eqref{eq:E+}. Previously, we explained using an analytical argument that $\Delta \mathcal{E}_+=0$. Here we will demonstrate this numerically using the trajectories from our simulations, obtained by solving the reduced LL equation with the total field $\mathcal{F}_i(x_i)$, including the $\lambda_0=100\,\si{\nano\metre}$ laser pulse as the external field. As the flux of the advanced field propagates in the opposite direction to the retarded field, we cannot simply integrate over the hemispherelike surface in Fig.~\ref{fig:detector}\,(a), but rather we must integrate over a spherelike surface as shown in Fig.~\ref{fig:detector}\,(b), in agreement with the integrated conservation law in Eq. (\ref{eq:Ttotalconserv}). Contributions of the retarded and advanced fields, propagating in opposite directions, are then expected to cancel. The spherelike surface is constructed by reflecting the hemispherelike surface in the $xy$ plane. Otherwise, the integration proceeds as before, with the same density of observation points. 

The magnitude of the first term in Eq.~\eqref{eq:E+}, which corresponds to the retarded field, is $0.0019\,\epsilon_0$ when integrated in the solid angle. In fact, this is simply one-quarter of $[\Delta\mathcal{E}]_{\text{incoh}}$ for $\lambda_0=100\,\si{\nano\metre}$, which we obtained earlier in Tab.~\ref{tab:energy_consv}. The magnitude of the second term of Eq.~\eqref{eq:E+}, which corresponds to the advanced field, is also $0.0019\,\epsilon_0$. Consequently, the difference between these terms in Eq.~\eqref{eq:E+} is zero, except for a small numerical error. Therefore, we have confirmed that $\Delta\mathcal{E}_+=0$.

\begin{acknowledgments}
\noindent
This article comprises part of the PhD work of Michael J. Quin, which was successfully defended at Heidelberg University on the 18th of October, 2023. The authors wish to thank John G. Kirk and Brian Reville for insightful discussions.

This material is based upon work supported by the Department of Energy [National Nuclear Security Administration] University of Rochester ``National Inertial Confinement Fusion Program'' under Award Number(s) DE-NA0004144.

This report was prepared as an account of work sponsored by an agency of the United States Government. Neither the United States Government nor any agency thereof, nor any of their employees, makes any warranty, express or implied, or assumes any legal liability or responsibility for the accuracy, completeness, or usefulness of any information, apparatus, product, or process disclosed, or represents that its use would not infringe privately owned rights. Reference herein to any specific commercial product, process, or service by trade name, trademark, manufacturer, or otherwise does not necessarily constitute or imply its endorsement, recommendation, or favoring by the United States Government or any agency thereof. The views and opinions of authors expressed herein do not necessarily state or reflect those of the United States Government or any agency thereof.
\end{acknowledgments}

\bibliography{bibliography}

\end{document}